\documentclass[12pt]{article}
\usepackage{astron}
\usepackage{color}
\usepackage{graphicx}

\def\degr{\hbox{$^\circ$}}

\def\fdg{\hbox{$.\!\!^\circ$}}

\def\farcs{\hbox{$.\!\!^{\prime\prime}$}}

\begin{document}

\title{67P/Churyumov-Gerasimenko -- potential
target for the Rosetta mission}
\author{Ma{\l}gorzata Kr{\'o}likowska \\
Space Research Centre of the Polish Academy of Sciences\\
Bartycka 18A, 00-716, Warsaw, Poland}

\maketitle

\begin{center}
email: mkr@cbk.waw.pl

Appeared in Acta Astronomica 53, pp 195-209 (2003)
\end{center}

\abstract{An influence of the non-gravitational effects on the
motion of short-period comet 67P/Chu\-ry\-u\-mov-Gerasimenko is
investigated. It was found that the normal component of the
non-gravitational force exceeds the transverse one and a model of
the motion including $A_1, A_2, A_3$ better fits the observations
than model neglecting $A_3$. Assuming asymmetry in $g(r)$ with
respect to the perihelion the large value of displacement $\tau$
was derived (about 34 days), and very small negative value of
transverse component $A_2$ was obtained. The models of rotating
non-spherical nucleus also suggest the large shift of light curve
with respect to perihelion ($\tau \ge 30$). The forced precession
model of 67P with $\tau = 34$~days gives a prolate spheroidal
shape of the rotating nucleus with axial ratio $R_b/R_a=1.16$,
rotational period to equatorial radius $P_{rot}/R_a=4.6\pm
1.4$~hrs/km, and torque factor $f_{tor}=3\cdot 10^5$~day/AU. The
much larger $\tau = 54$~days gives distinctly prolate shape of
nucleus with axial ratio $R_b/R_a=1.71$. The orientation of spin
axis of the nucleus and its evolution are presented. The past and
the future dynamical evolution of comet 67P is also widely
discussed.}


\section {Introduction}

Comet 67P/Churyumov-Gerasimenko was discovered by Klim Churymov on
photographs of 32P/Comas Sola taken by Svetlana Gerasimenko on
September 1969 at Alma Ata. From that moment 67P has been detected
during its all six returns (Marsden and Williams 2001, Rocher
2003). Now it is still extensively observed during its sixth
apparition and it will be potentially observable up to beginning
of 2004.

67P/Churyumov-Gerasimenko is unusually active for a short-period
comet with the period of 6.6~yr. In the current apparition the
comet passed the perihelion on 18 August 2002 and peaked at around
magnitude 12, although  an outburst of approximately 2 magnitudes
at perihelion has been reported. Similar phenomenon was seen in
the previous return but with a slightly lower amplitude. On both
occasions the rise in the light curve was rapid; the light-curves
are presented by Yoshida (2003) at WEB pages. It appears that the
1996 outburst came a few days before perihelion passage, whereas
the 2002 event was centered exactly on perihelion. Besides, in the
last perihelion passage the tail has been extended over 10
arcminutes,  and seven months after perihelion it still was very
well developed.

Contrary to such photometric activity no major orbital changes
occurred since its discovery in 1969. However, the comet has
rather unusual history. During 19$^{th}$ and 20$^{th}$ centuries
prior to 1959 its perihelion distance varied from about 2.5~AU to
2.9~AU, and this is the reason why the comet was unobservable from
the Earth at that time. In February 1959 the close approach of the
comet to Jupiter to within 0.052~AU occurred. This event caused
considerable orbital changes: the perihelion distance has been
reduced from 2.74~AU to 1.28~AU, the eccentricity increased from
0.36 to 0.63 and orbital period shortened from 8.97~yrs to
6.55~yrs. As a result the comet was discovered in its second
return to perihelion after the close encounter with Jupiter.
Hence, the unusual comet activity detected in the last apparitions
could be a result of this remarkably reduction of perihelion
distance taking place not long ago.

67P/Churyumov-Gerasimenko has just been selected as the new target
for the Rosetta mission after the failure to launch the probe  for
an encounter with 46P/Wirtanen. This inspired me to investigate
the non-gravitational effects in the motion of this comet.

\section{Observational material and the method of calculations}

The present investigations are based on the archive observations
available at the Minor Planet Center (Cambridge, USA). The whole
observational material contains 1207 observations covering the
time period from 1969 September 9 to 2003 March 13. The
observations were selected according to the objective criteria
elaborated by Bielicki and Sitarski (1991) for each of six
apparitions separately. Finally, 2338 residuals were used for the
orbit improvement.

The non-gravitational equations of cometary motion have been
integrated numerically using recurrent power series method
(Sitarski 1989, 2002) taking into account the perturbations by all
the nine planets. All numerical calculations presented here are
based on the Warsaw numerical ephemeris DE405/WAW of the Solar
System, consistent with high accuracy with the JPL ephemeris DE405
(Sitarski 2002). The standard epoch of 2003~Dec.~27 was accepted
in all the calculations as the starting epoch of integration.

\section{Models with constant non-gravitational parameters}

To estimate the non-gravitational force acting on the rotating
cometary nucleus with sublimating water from its surface the
standard Marsden method (Marsden et al. 1973) was used. This
formalism assumes that the three components of a non-gravitational
acceleration have a form:
\begin{equation}
F_{\rm i}   =  A_{\rm i} \cdot g(r), \qquad A_{\rm i} = {\rm
~const~~~for} \quad {\rm i}=1,2,3,
\end{equation}
\noindent where $F_1,F_2,F_3$ represent the radial, transverse and
normal components of the non-gravitational acceleration,
respectively. The function $g(r)$ simulates the ice sublimation
rate as a function of the heliocentric distance $r$:
\begin{equation} g(r) = \alpha \left( r/r_o \right) ^{-m} \left[ 1
+\left( r/r_o \right) ^{n} \right] ^{-k},
\end{equation}
\noindent where the exponential coefficients $m,n$ and $k$ are
equal to $2.15$, $5.093$, and $4.6142$, respectively. The
normalization constant $\alpha =0.1113$ gives $g(1$~AU$)=1$; the
scale distance $r_0=2.808$~AU.

To generalize Eqs.~1-2 of the non-gravitational effects to
asymmetric case in respect to perihelion we simply substitute
$g(r\prime )$ instead of $g(r)$, where $r\prime = r(t-\tau)$, and
the time shift $\tau$ represents the time displacement of the
maximum of the function $g(r)$ with respect to the perihelion.

For whole time interval the constant values of $A_1, A_2$ and
$A_3$ (and eventually $\tau$ in the asymmetric case) were
calculated along with the six corrections to the orbital elements.
The results are given in Table~1 as Model Ia and Model~Ib, for the
symmetric and asymmetric case, respectively. The solution of using
the asymmetric non-gravitational acceleration model do not
significantly decreases the {\it rms} in comparison to symmetric
model (see also Yeomans and Chodas (1989)), however the
photometric observations give arguments for clear asymmetry in the
light curves of 67P. The derived time shift $\tau=34.3$ is in an
excellent agreement with the visual light curve of 67P obtained by
Morris (Hanner et al. 1985). He show that the light curve of the
comet during its 1982-83 apparition reached a brightness maximum
approximately 35 days after perihelion. The result of 34.3~days
for the time shift is also in a good agreement with the light
curves of the comet presented in WEB page by Kidger (2003). It
seems that photometric and positional observations of 67P
independently provide consistent determination of $\tau$. I
repeated the calculations for shorter arc of 1982~May~31 --
2003~Mar.~13, i.e. only for last four apparitions. Surprisingly,
quite the same time shift of $\tau=34.9$~days was obtained.

\begin{table}
\caption{Non-gravitational parameters and orbital elements for the
67P/Churyumov-Gerasimenko derived from all positional observations
(six apparitions). Non-gravitational parameters $A_1, A_2, A_3$
are given in units of $10^{-8}$AU$\cdot$day$^{-2}$. Angular
elements $\omega$, $\Omega$, $i$ are referred to Equinox J2000.0
(Epoch: 20031227). Numbers in parentheses denote uncertainties:
$0.63175088(7) \equiv 0.63175088 \pm  0.00000007$}
\begin{center}
{\setlength{\tabcolsep}{2.0mm} {\small \vspace{0.10cm}
\begin{tabular}{ccc}
  \hline
           & Model Ia                 & Model Ib \\
   && \\
   \hline
   && \\
  $A_1$    & 0.054440$\pm$0.002665    & 0.088327$\pm$0.003598  \\
  $A_2$    & 0.0098084$\pm$0.0000173  & $-0.0013637\pm 0.0009429$  \\
  $A_3$    & 0.030187$\pm$0.002189    & 0.033855$\pm$0.002152  \\
  $\tau$   & ---                      & 34.314$\pm$2.128 \\
   && \\
  $T$        & 20020818.28695(7)      & 20020818.28685(6) \\
  $q$        & 1.29064789(25)         & 1.29064249(15)    \\
  $e$        & 0.63175088(7)          & 0.63175242(4)     \\
  $\omega$   & 11\degr 40976(8)       & 11\degr 40848(7)  \\
  $\Omega$   & 50\degr 92865(7)       & 50\degr 92965(7)  \\
  $i$        & 7\degr 12415(1)        & 7\degr 12413(1)   \\
  $rms$      & 1\farcs 16             & 1\farcs 12        \\
  \hline
\end{tabular}
}}
\end{center}
\end{table}

One can see that normal component of non-gravitational
acceleration is significantly greater than transverse component
for both models. To compare our analysis of the non-gravitational
effects with those published by other authors who neglected the
normal component, I have repeated all calculation assuming
$A_3=0$; the results are summarized in Table~2. This model fits
the observations with the $rms$ larger than Model~Ia by about
0\farcs 04.

\begin{table}
\caption{Non-gravitational parameters $A_1$ and $A_2$ obtained
with assumption that normal component $A_3$ is equal zero}
\label{tab2} \begin{center} {\setlength{\tabcolsep}{1.0mm} {\small
\vspace{0.10cm}
\begin{tabular}{ccc} \hline
 Arc of observations  & $A_1$ & References \\
    No of apparitions & $A_2$ &            \\
    $rms$             & in units of $10^{-8}$~AU/day$^2$  & \\
    && \\ \hline
    && \\
   1969 -- 1988 (229 obs.) & $A_1=+0.069\pm 0.020$  & Chodas \& Yeomans (1989)  \\
  4 app; $rms=1$\farcs 34  & $A_2=+0.010\pm 0.001$  & \\
    && \\
  1969 -- 1997 (474 obs.) & $A_1=+0.07$             & MPC 34423  \\
  5 app; $rms=1$\farcs 0  & $A_2=+0.0099$           & \\
    && \\
  1975 -- 1997 (424 obs.) & $A_1=+0.07\pm 0.007$    & Muraoka   \\
  4 app; $rms=0$\farcs 85 & $A_2=+0.0094\pm 0.0002$ & www.aerith.net\\
    && \\
  1975 09 09 -- 2003 01 06   & $A_1=(+0.05037\pm 0.00226)$ & Cometary Notes\\
  5 app (747 obs.)           & $A_2=(+0.00936\pm 0.00002)$ &  of Bureau des longi-\\
  $rms=0$\farcs 70           &                             & tudes (note no 29)    \\
   && \\
  1969 08 08 -- 2003 03 13   & $A_1=(+0.04954\pm 0.00275)$   & \\
  6 app (1207 obs.)          & $A_2=(+0.009786\pm 0.000018)$ & present  \\
  $rms=1$\farcs 20           && calculations  \\
  \hline
\end{tabular}
  }}
\end{center}
\end{table}

\section{Rotating cometary nucleus and the forced precession model}

\begin{table}
\caption{Physical parameters for rotating cometary nucleus and
orbital elements linking  all positional observations (six
apparitions) of 67P/Churyumov-Gerasimenko. Angular elements
$\omega$, $\Omega$, $i$ are referred to Equinox J2000.0.
Non-gravitational parameter $A$ is given in units of
$10^{-8}$~AU/day$^2$, the precession factor $f_{\rm p}$ is in
units of $10^6$~day/AU, time shift $\tau $, is in days. Subscript
'0' in $I_0$ and $\phi _0$ denotes the values on the starting
epoch of integration (Epoch: 20031227)} \label{tab3}
\begin{center}
{\setlength{\tabcolsep}{2.0mm} {\small \vspace{0.10cm}
\begin{tabular}{cccc}
  \hline
  & Model IIa & Model IIb   & Model III          \\
  & \multicolumn{2}{c}{}  & Forced precession  \\
  & \multicolumn{2}{c}{}  & model              \\ \hline
  &&& \\
    $A$        & $0.064173\pm 0.003258$          & $0.096424\pm 0.003519$    & 0.10345$\pm$0.00560      \\
  $\eta$     & 35\degr 90$\pm$3\degr 35        & 27\degr 56$\pm$1\degr 90  & 28\degr 29$\pm$2\degr 17 \\
  $I_0$      & 72\degr 49$\pm$0\degr 99        & 90\degr 38$\pm$0\degr 91  & 88\degr 19$\pm$0\degr 76 \\
  $\phi _0$  & 333\degr 98$\pm$6\degr 88       &317\degr 25$\pm$5\degr 01  & 315\degr 24$\pm$5\degr 63\\
  $f_p$      & ---                             & ---                       & $-1.849\pm 1.360$        \\
  $s$        & ---                             & ---                       & $-0.1613\pm 0.0685$      \\
  $\tau$     & ---                             & 34.314                    & 34.314                   \\
  &&& \\
  $T$        & 20020818.28696(7)     & 20020818.28686(7)    & 20020818.28668(7)    \\
  $q$        & 1.29064840(25)        & 1.29064313(25)       & 1.29064511(25)       \\
  $e$        & 0.63175074(7)         & 0.63175225(7)        & 0.63175157(7)        \\
  $\omega$   & 11\degr 40936(8)      & 11\degr 40783(8)     & 11\degr 40785(8)     \\
  $\Omega$   & 50\degr 92908(7)      & 50\degr 93033(7)     & 50\degr 93037(7)     \\
  $i$        & 7\degr 12413(1)       & 7\degr 12410(1)      & 7\degr 12410(1)      \\
  $rms$      & 1\farcs 16            & 1\farcs 12           & 1\farcs 11        \\
  \hline
\end{tabular}
}}
\end{center}
\end{table}

The non-gravitational parameters for the model of the rotating
spherical nucleus were also determined. In such model three
parameters $A_1, A_2$ and $A_3$ are now the functions of time by
relations: $A_i=A\cdot C_i(t)$, $i=1,2,3$, where $C_i(t)$ are
direction cosines for the non-gravitational force acting on the
rotating cometary nucleus (Kr\o likowska et al. 1998). Three next
non-gravitational parameters describing the model of the rotating
cometary nucleus are angular parameters: $\eta$ -- the lag angle
of the maximum outgassing behind subsolar meridian, $I$ --
equatorial obliquity and $\phi$ -- cometocentric solar longitude
at perihelion. The values of four parameters, $A, \eta, I$ and
$\phi$, are presented in Table~3 as Model IIa, and IIb for the
symmetric and asymmetric cases, respectively. In the Model~IIb the
value of time shift $\tau$ was taken from Model Ib.

\begin{figure}
{\centering
 \includegraphics[width=12cm,angle=0.0]{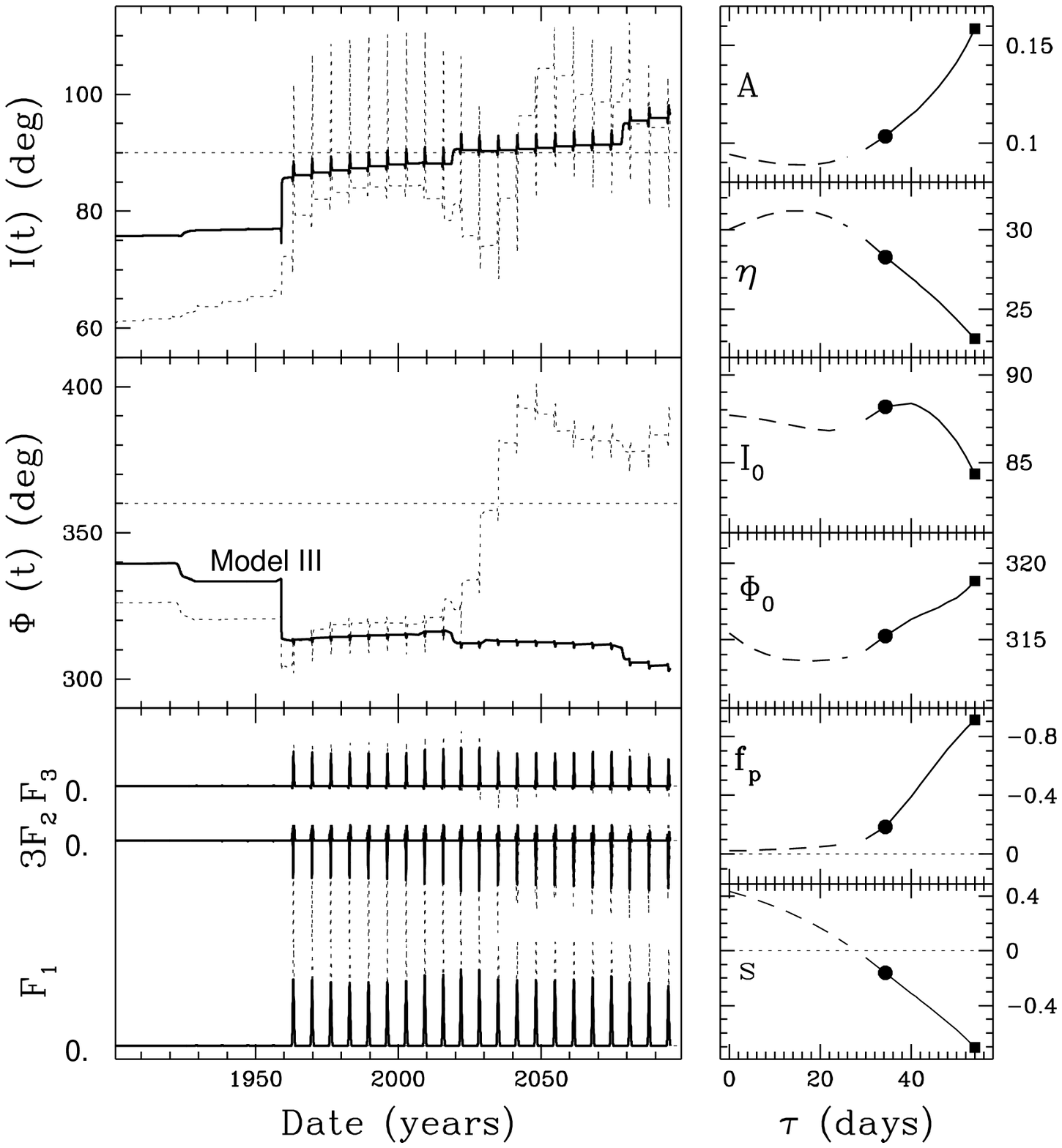}
\caption{{\bf Left side:} Temporal variation of angle $I, \phi$
and components $F_1, F_2, F_3$ of the non-gravitational force due
to the spin axis precession of the nucleus of
67P/Churyumov-Gerasimenko. Dashed horizontal line on the upper
panel divides models with prograde rotation (consistent with the
sense of the cometary orbit around the sun, $I < 90^0$) from
models with retrograde rotation ($I > 90^0$); dashed horizontal
line on the middle panel indicates changes of configuration of
poles in the perihelia. Thick, solid curves show the evolution of
$I, \phi ,F_1, F_2, F_3$ for Model~III represented by black dots
in the right-side panel (see also Table~3), and thin, dashed
curves give the respective evolution for the model represented by
black squares in the right-side panel.
\newline \noindent {\bf Right side:} Family of
forced precession models parameterized by time shift $\tau$ and
all fit the observations with the $rms=$1\farcs 11. From top to
bottom: $A$ -- the non-gravitational parameter given in units of
$10^{-8}$~AU/day$^2$, $\eta$ -- the lag angle , $I_0$ -- the
equatorial obliquity of the spin axis relative to the orbital
plane, $\phi_0$ -- the cometocentric solar longitude at
perihelion, $f_p$ -- the precession factor given in units of
$10^7$~day/AU, and $s$ -- the oblateness of the nucleus. Subscript
'0' in $I_0$ and $\phi _0$ denotes the values on the starting
epoch of integration (Epoch: 20031227). The dashed parts indicate
that models are nonphysical ($f_{tor} = f_p \cdot s < 0$). The
black dots show the position of Model~III (see Table~III) and
black squares -- the forced precession model with the time shift
$\tau = 54.0$.
 }}
\end {figure}

An assumption of the flattened nucleus represents the next step
towards the more realistic cometary models. In this case, the
forced precession of the spin axis could arise due to a torque if
a vector of the jet force does not pass through the center of the
nonspherical nucleus. The precession rate is a function of the
nucleus orientation, the lag angle ,$\eta$, the modulus of the
reactive force ,$A$, the nucleus oblateness ,$s$, and the
precession factor ,$f_p$, which depends of the rotation period and
nucleus size. In such model the six parameters are derived: $A,
\eta, I, \phi, s$ and $f_p$ (Kr\'o likowska et al. 1998).
Unfortunately the seventh parameter $\tau$ was impossible to
obtained from the observational data. Thus, the right panel in
Fig.~1 shows family of forced precession models parameterized by
the time shift $\tau$. Each model from the family fits the
observational data with the same $rms=$1\farcs 11. Since the
precession factor $f_p$ is related to the torque factor, $f_{tor}$
introduced by Sekanina (1984) by $f_p = s\cdot f_{tor}$, $f_p$ and
$s$ should have both positive or both negative values. For $\tau <
30$~days the negative values of $f_p$ and positive values of $s$
were obtained. Therefore, the symmetric forced precession models
as well as forced precession models with negative or small
positive value of $\tau$ were excluded. It turns out that only the
models with significant positive time shift are allowed. Table~3
shows the fully consistent forced precession model (Model III)
where the value of time shift $\tau$ was taken from Model~Ib. The
negative value of $s$ found in this forced precession model
suggests that nucleus of Comet 67P/Churyumov-Gerasimenko has a
prolate spheroidal shape with axial ratio $R_b/R_a=1-s=1.16$,
where, $R_b$ and $R_a$ are the polar and equatorial radii,
respectively. The motion of the cometary rotation axis represented
by angles $I$ and $\phi$ is pointed by thick, solid curves in the
left panel of Fig.~1. Qualitative variations of the
non-gravitational force components acting on the comet during its
successive returns to the sun are also shown. One can see that the
significant reduction of the perihelion distance in 1959 caused
significant increase of the non-gravitational force. After 1959
the time variations of components of non-gravitational force are
rather regular. This model is also attractive since its very
moderate values of $f_p$ and $s$ which give small value of torque
factor $f_{tor}= 3 \cdot 10^5$~day$\cdot$AU$^{-1}$. There are also
possible models with larger $\tau$ and one of them with
$\tau=54$~days is visualized in Fig.~1 by thin, dotted curve for
comparison with Model~III. This model (black squares in the left
panel of Fig~1) is characterized by distinctly prolate shape of
nucleus with axial ratio $R_b/R_a=1-s=1.71$ and torque factor 20
times greater than that in Model~III ($f_{tor}= 6.4 \cdot
10^6$~day$\cdot$AU$^{-1}$).

\section{Orbital evolution}

\begin{figure}
{\centering
 \includegraphics[width=14cm,angle=0.0]{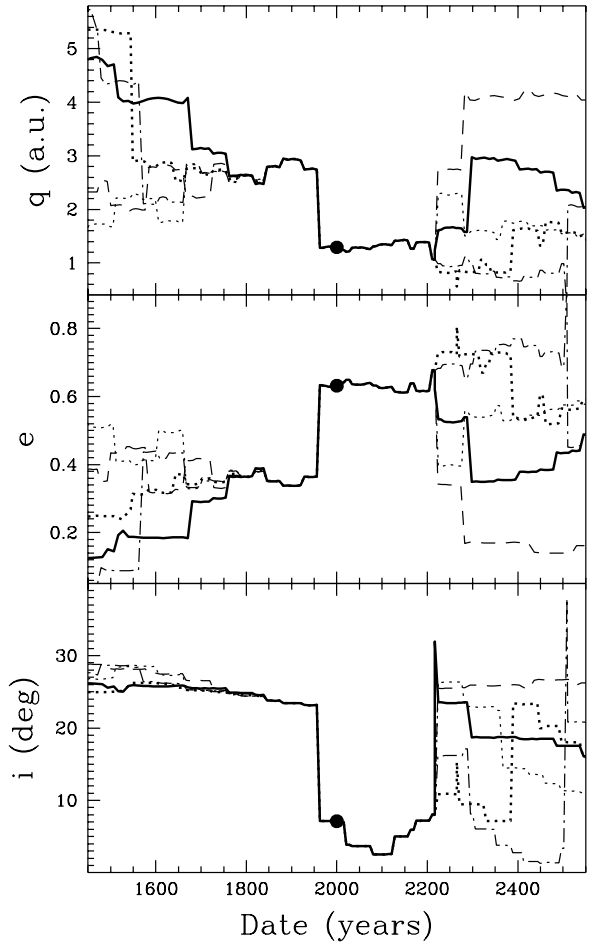}
\caption{Time evolution of the orbital elements $q$, $e$ and $i$
of 67P Churyumov-Gerasimenko. The present orbit of the comet
determined from whole observational data (1969~Sept.~9 --
2003~Mar.~13) was integrated for over 500 yrs back and forward
(starting point -- 2003~Dec~27 -- is indicated by black dots) and
its non-gravitational evolution is shown with:
\newline solid, thick lines -- for evolution correspond to Model~Ia
\newline dotted, thick lines -- for evolution correspond to Model~Ib
\newline dashed lines -- for evolution correspond to Model~IIa
\newline dotted-dashed lines -- for evolution correspond to Model~IIb
\newline dotted lines -- for evolution correspond to Model~III
 }}
\end {figure}

\begin{figure}
{\centering
 \includegraphics[width=14cm,angle=0.0]{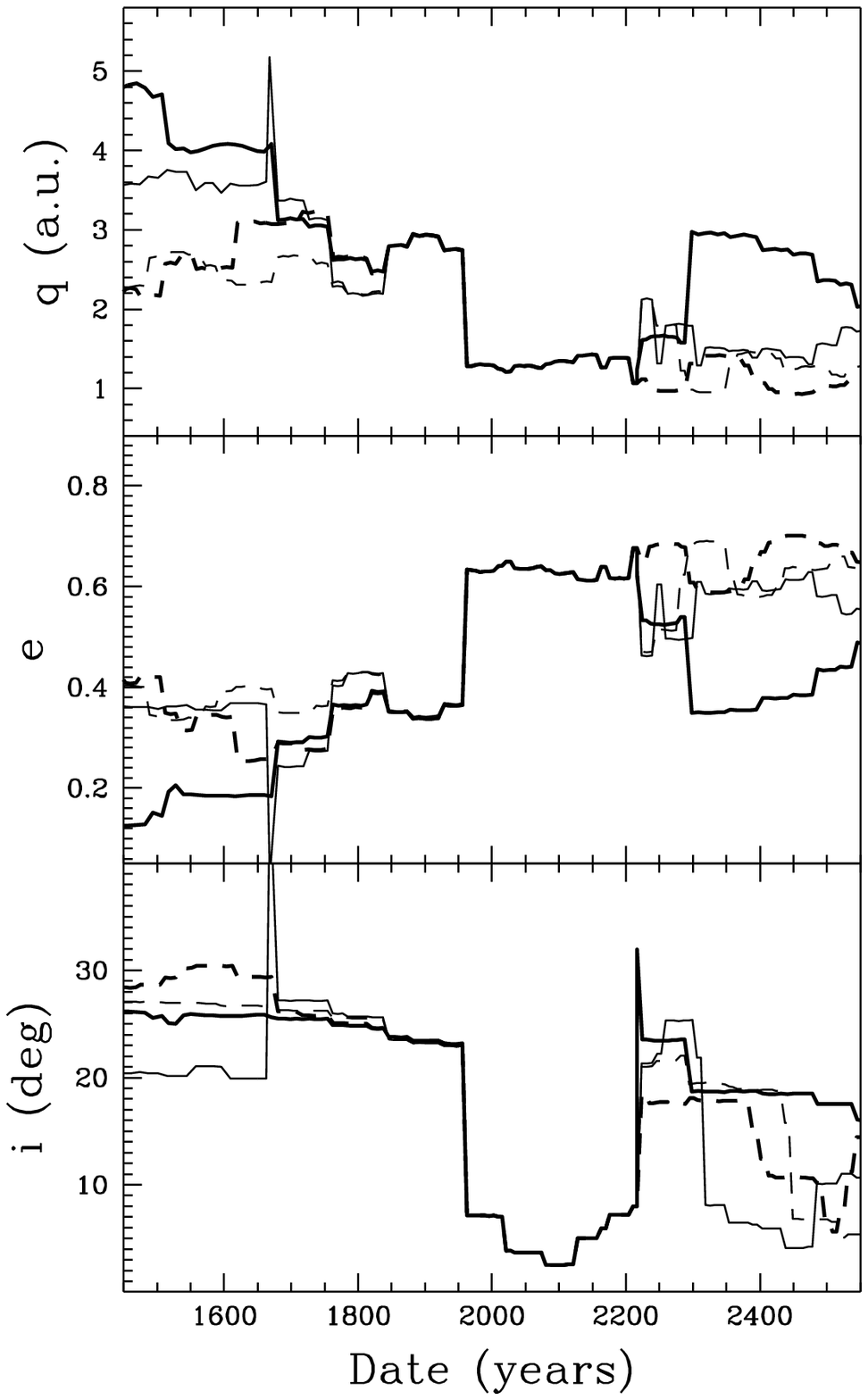}
\caption{Time evolution of the orbital elements $q$, $e$ and $i$
of 67P Churyumov-Gerasimenko for Model~Ia (thick and thin solid
lines) and model with $A_3=0$ (thick and thin dotted lines). The
thick lines represent the non-gravitational evolution, the thin
lines show pure gravitational evolution which starts from the same
orbital elements as in the non-gravitational case. In the model
with $A_3=0$ the value of $A_1=(+0.04954\pm 0.00275)$ and
$A_2=(+0.009786\pm 0.000018)$ in units of $10^{-8}$~AU/day$^2$ are
derived from observations (see also Table~2).
 }}
\end {figure}

Studying the evolution of short-period comets belonging to the
jupiter-family comets we should confine to short-term numerical
integrations, say up to one thousand years in the past as well as
in the future. Accordingly, the dynamical evolution of Comet 67P
was followed up to one milenium from the starting epoch of
integration (2003~Dec.~27). Calculations were performed for all
five models discussed in the previous sections (Tables~1~\& 2).
Fig.~2 shows the non-gravitational evolution of the perihelion
distance, $q$, the eccentricity, $e$, and the inclination, $i$.
Comparison between non-gravitational evolution and evolution
without non-gravitational effects is presented in Fig.~3 for
starting orbital elements taken from Model~Ia (solid lines) and
the model with the assumed normal non-gravitational parameter
$A_3=0$ (dotted lines), respectively.

In the case of Comet Churyumov-Gerasimenko the very close approach
to Jupiter in 2020 makes future orbital predictions for the longer
time than two century uncertain. Before 2020, six encounters with
Jupiter will take place (in 2018, 2078, 2125, 2161, 2172, and 2209
for all five models, see Fig.~4), however without any spectacular
orbital changes. After close encounter with Jupiter in 2020 the
future evolution will be unpredictable, but some evolution
similarities are visible. In all five models many approaches to
Jupiter are expected between 2220--2500 (Fig.~4) with at least one
closer than 0.1~AU. During last five centuries significantly less
approaches to Jupiter happened than will occur in the next 500
years. In February 1959 -- as was mentioned before -- the comet
approached Jupiter to within 0.052~AU what caused considerable
changes in orbital elements, especially in the perihelion
distance: it was reduced from 2.74~AU to 1.28~AU, and, in the
consequence, the comet was discovered soon after that. Fig.~3
clearly shows that in the same event the eccentricity increased
from 0.26 to 0.63 and the orbital inclination decreased from
23\fdg 2 to 7\fdg 2. Past encounters with Jupiter occurred in
roughly one century time intervals. Up to now many researches
conclude that 67P is a comparatively recent visitor to the inner
solar system having had $q=4.0$~AU prior to 1840 and $q>2.75$~AU
until a Jupiter encounter in 1959. However, Figs.~2--3 show that
the evolution before 1700 is highly uncertain.

\begin{figure}
{\centering
\includegraphics[width=12cm]{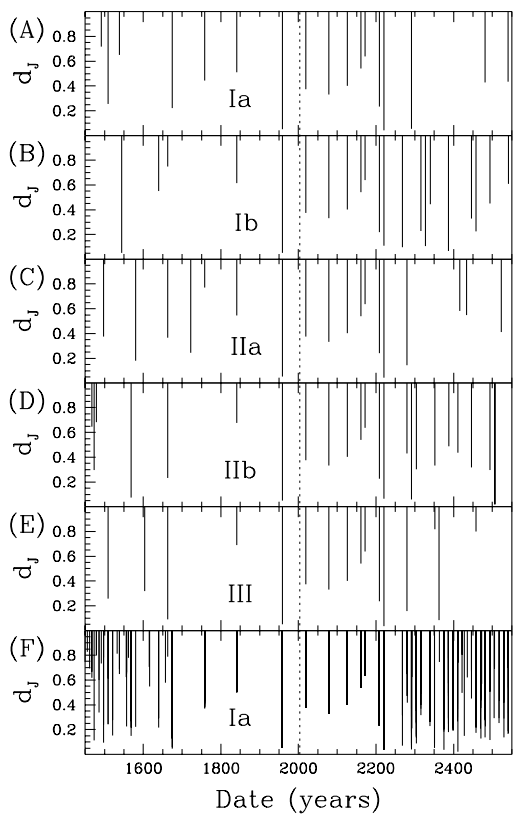}
\caption{Distribution of all close approaches of the comet to
Jupiter which appeared during evolution of the starting orbit in
the Models Ia (panel (A)), Ib (panel (B)), IIa (panel (C)), IIb
(panel (D)), III (panel (E)), respectively. Panel (F) presents
cumulative distribution of all close encounters with Jupiter which
occurred in evolution of 20 clones of orbit constructed from the
Model~Ia. The evolution was performed backwards and forwards up to
500 yrs. Starting moment of integrations is shown by dotted
vertical line. The y-axis ($d_J$) shows the depths of individual
close encounters of the comet with Jupiter (closer than 0.8 AU).}}
\end{figure}

\begin{figure}
{\centering
 \includegraphics[width=13cm]{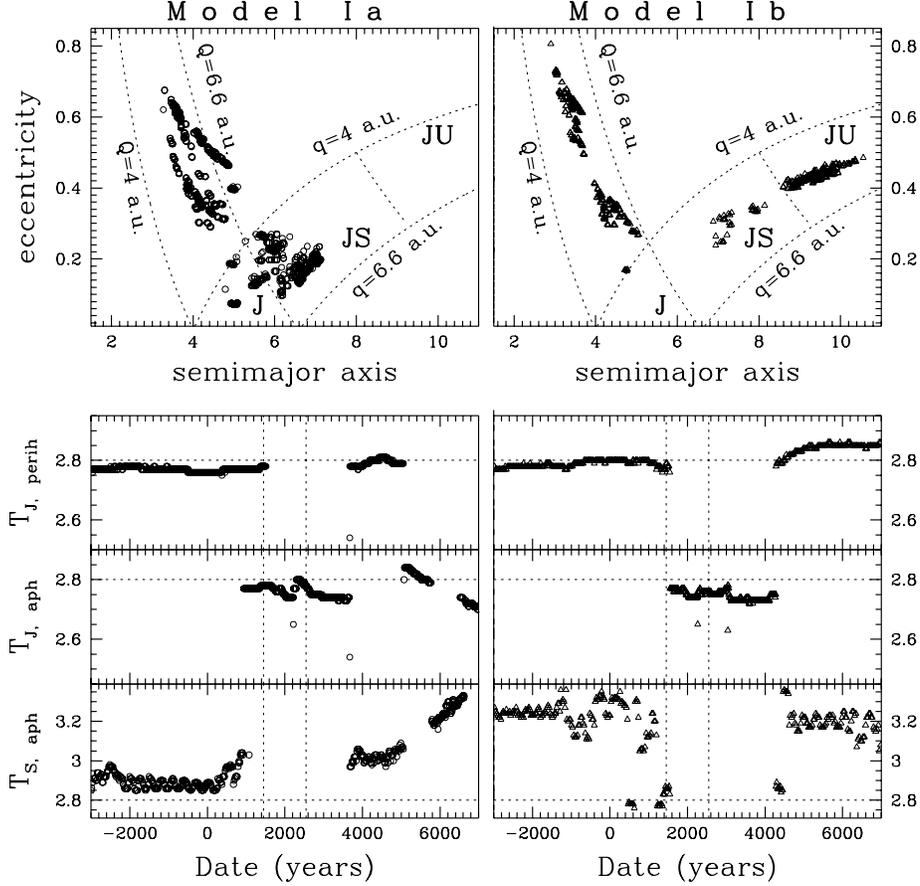}
\caption{{\bf Upper panels:} Plot of eccentricity versus semimajor
axis for evolution of comet 67P according to Model~Ia (the
left-hand side) and Model~Ib (the right-hand side), respectively.
The dashed thin curves marks the boundaries of the aphelion or
perihelion zones controlled by Jupiter. Jupiter's zone of control
is taken as a three times Hill radii. In the zone of
4~AU~$<q<$~6.6~AU to the right of line $Q>$~4~AU the regions
belonging to SP comet's categories: J (objects for which both
perihelion and aphelion are under Jupiter's control), JS
(perihelion is under Jupiter's and the aphelion under Saturn's
control), and JU (perihelion -- as previously, and the aphelion
under Uranus's control) (Horner et al. 2003). The evolution was
performed backwards and forwards up to 5 kyr.
\newline \noindent {\bf Lower panels:} Changes of Tisserand
parameter, T$_{J, perih}$ (T$_{J, aph}$, T$_{S, aph}$) during time
interval in which the perihelion (aphelion) falls in the Jupiter's
(or Saturn's) zone of control. The horizontal dashed lines show
the boundaries differentiate between SP comets using the Tisserand
parameter. After Horner et al. (2003) two of this four-fold
division are denoted by III (third class having 2.5~AU~$ \ge T_J >
$~2.8~AU) and IV (fourth class having $T_J \ge 2.8$~AU) correspond
to loosely bound Jupiter-family comets and tightly bound
Jupiter-family comets, respectively.
 }}
\end {figure}

\begin{figure}
{\centering
 \includegraphics[width=14cm]{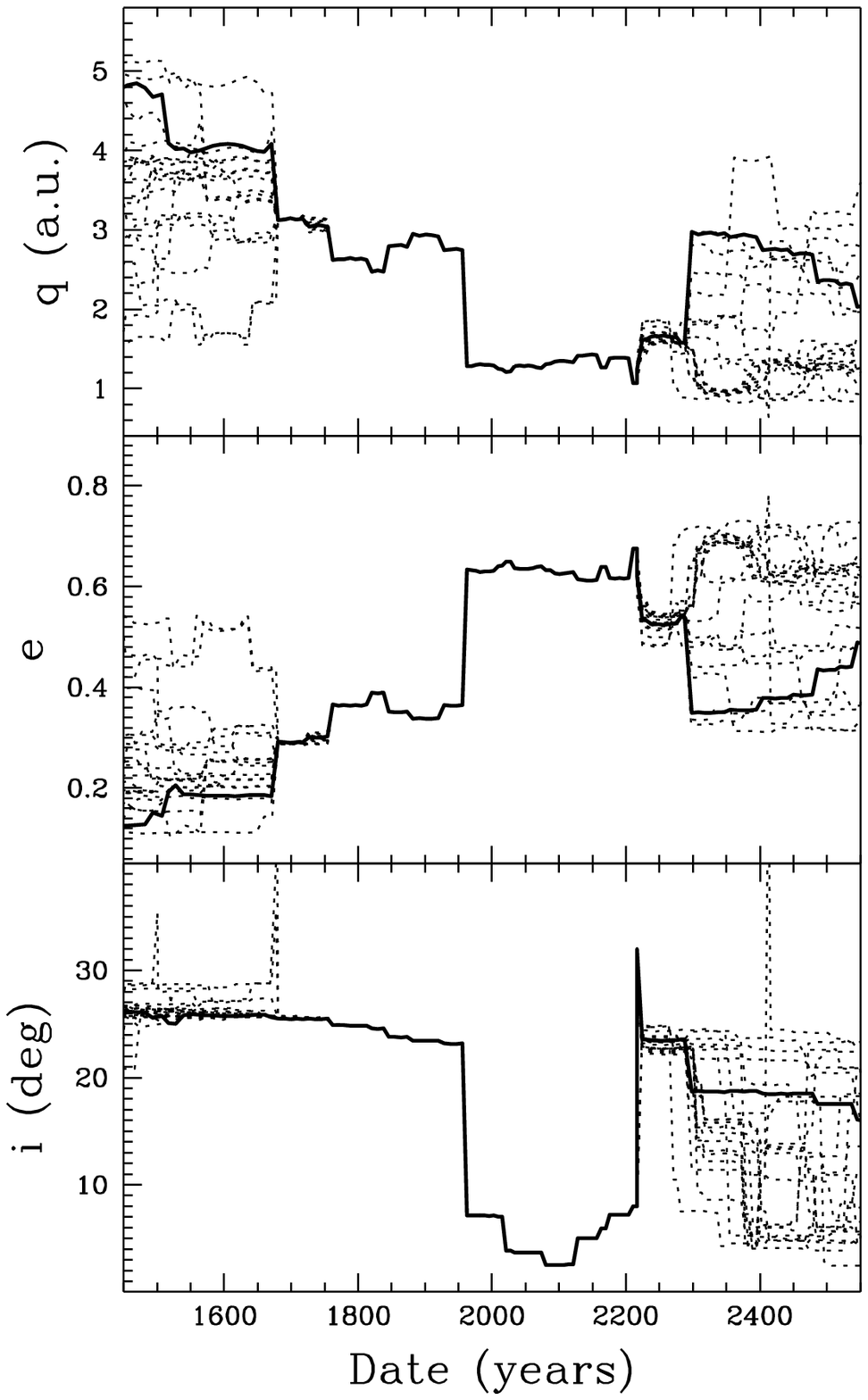}
\caption{Time evolution of the orbital elements $q$, $e$ and $i$
of 20 randomly selected orbits of 67P/Churyumov-Gerasimenko.  The
evolution was performed backwards and forwards up to 500 yrs from
the starting moment of integrations (2003~Dec~27). The nominal,
starting orbit of the comet corresponds to Model~Ia and its
dynamical evolution is given by solid, thick curve.
 }}
\end {figure}

   Keeping this in mind let me only speculate about past and future
evolution extending over $\pm$5 kyr. The evolutionary calculations
were performed to demonstrate how Jupiter controls the evolution
of the Comet Churyumov-Gerasimenko. To do this the classification
scheme proposed by Horner (2003) was applied for evolutionary
non-gravitational calculations according to Model~Ia (the
left-hand side of Fig.~5) and Model~Ib (the right-hand side of
Fig.5). The evolution in Model~Ia is different in details from the
evolution in Model~Ib. Starting at -3000 BC the cometary orbits
were placed in the JS region in the upper panel of Fig~5 and
started to evolve to the left (Model~Ia) or firstly to the right
to JU region and back to the left (Model~Ib). During the dynamical
evolution the perihelion distance of 67P being under the Jupiter
control (4.0~AU$<q<$6.6~AU) kept its value within
5.0~AU~$<q<5.8$~AU, (Model~Ia) or slowly evolved from $\sim$4.0~AU
to $\sim$5.0~AU (Model~Ib). The Tisserand parameter $T_J$ was
around the boundary value of 2.8 which divides loosely bound
Jupiter-family comets (class III in Horner (2003)) from tightly
bound JFC (class IV therein). Thus, during past evolution prior to
comet's discovery the perihelion was controlled by Jupiter and
aphelion was placed in the Saturn zone of control (Fig.~5). In the
Model~Ia about 950~AD also the aphelion started to be under
Jupiter's zone of control and prior to $\sim$~1500~AD both
perihelion and aphelion were under Jupiter's control (J class, see
Fig.~5). Then the cometary orbit with $e<0.2$ and semimajor axis
$a\simeq 5$ is placed in the lowest part of upper panel in the
Fig.~5. After 1500~AD the aphelion of the comet become under
Jupiter control and will be under Jupiter control prior to about
4000 AC. The lower panel of Fig.~5 shows that in almost whole time
interval considered in the Figs~2--4 the aphelion of 67P is under
Jupiter's control, e.g. the aphelion distance falls between
4.0~AU$<Q<$6.6~AU (Models~Iab). Future evolution shows some
similarities to the past evolution. Once again the perihelion
starts to be in the Jupiter's zone of control, and perihelion --
within Saturn's zone of control (Model~Iab) and next within Uran's
zone of control (Model~Ib).

\begin{table*}
\caption{Dispersion of non-gravitational parameters and orbital
elements derived for 20 randomly selected orbits of
67P/Churyumov-Gerasimenko (Epoch: 20031227; Equinox: J2000.0)}
\begin{center}
{\setlength{\tabcolsep}{1.0mm} {\small \vspace{0.10cm}
\begin{tabular}{rrrrrr}
  \hline
     $A_1$    &    $A_2$   &    $A_3$   & & & \\
 \hline
  &&&&& \\
     0.05444  &   0.009808 &   0.03019  & & & \\
  $+$0.00530  &$+$0.000022 &$+$0.00254  & & & \\
  $-$0.00626  &$-$0.000046 &$-$0.00546  & & & \\
  &&&&& \\
 \hline
   $T$ & $q$ & $e$ & $\omega$ & $\Omega$ & $i$ \\
  \hline
  &&&&& \\
 20020818.28695 &    1.29064789 &    0.63175088 &  11.40976 &  50.92865 &   7.12415 \\
     $+$0.00011 & $+$0.00000041 & $+$0.00000012 &$+$0.00012 &$+$0.00015 &$+$0.00001 \\
     $-$0.00014 & $+$0.00000043 & $+$0.00000013 &$+$0.00015 &$+$0.00012 &$+$0.00003 \\
 &&&&& \\ \hline
\end{tabular}
}}
\end{center}
\end{table*}

Additionally, the statistical approach was used for the study of
the past and future motion of comet 67P. The sample of 20 clones
of nominal orbit were constructed according to the method
described by Sitarski (1998) for the standard symmetric model
(Model~Ia) with constant non-gravitational parameters $A_1, A_2$
and $A_3$. Sitarski's procedure allows to derive the set of
randomly selected orbits (clones) which all fit the observations
almost with the same {\it rms} as nominal orbit. The range of six
orbital elements taken as starting orbit to dynamical calculations
are given in Table~4. Next, each randomly selected orbit was
integrated backwards and forwards up to 1 kyr. The differences in
past and future evolution of $q$, $e$ and $i$ are clearly visible
in Fig.~6. The evolution is well defined in the period of [-300;
+250] years, outside this time-interval dynamical behaviour starts
to be chaotic.

Comet Churyumov-Gerasimenko belongs to so called Near Earth Comets
(NEC) (Baalke 2003). In the present dynamical calculations
covering the time period 1500--2500 the close encounters with the
Earth to within 0.2~AU were analyzed. Such events were found only
for the Model~IIb among five nominal orbits described in
Table~1~\&~2, and for 10 of 20 randomly selected orbits in the
Model~Ia. The first close encounter with the Earth occurs in 2239
(to within 0.045~AU, Model IIb), i.e. after remarkably close
encounter with Jupiter in 2220. After that the series of close
approaches to the Earth were detected for this nominal orbit
(comet runs inside the earth's orbit). For randomly selected orbit
the closest approaches to within 0.019~AU were found among
cumulative number of 19 encounters with the Earth to within
0.2~AU.

\section{Conclusions}

Considering the results obtained from different models of the
non-gravitational motion of jupiter-family comet
67P/Churyumov-Gerasimenko, the principal conclusions from this
study are following.

\begin{enumerate}
    \item The non-gravitational effects detected in the comet
    Churyumov-Gerasimenko motion seem to be small and stable
    during all six apparitions despite different outbursts
    observed in the light-curves of this comet.
    \item The normal component of the non-gravitational force
    exceeds the transverse one and the model of motion including
    $A_1, A_2, A_3$ better fits the observations of 67P than the model
    with neglecting $A_3$ (Tables~1~\& 2).
    \item Investigation of the non-gravitational motion of the
    rotating cometary nucleus of 67P indicates a large value of the time
    shift $\tau > 30$~days which measures the displacement of
    maximum of $g(r)$ with respect to perihelion. Thus the value
    of $\tau =34$~days independently derived from the model with
    constant parameters $A_1, A_2, A_3$ is consistent with the
    forced precession model.
    \item The forced precession model of 67P with assumed $\tau=34$~days
    gives a prolate spheroidal shape of the rotating nucleus with axial
    ratio $R_b/R_a=1.16$, and $P_{rot}/R_a=4.6\pm 1.4$~h/km.
    This value of $P_{rot}/R_a$ lies within the region occupied by comets with known sizes
    and rotational periods (Fig.~5 in Kr{\'o}likowska et al. (2001)).
    \item Dynamical evolution of 67P is well defined to about
    $\sim$~250--300 years backward and forwards in time.
    \item Considering the results of short-time integrations
    (within the interval of $\pm$~1~kyr) of 20 randomly
    selected orbits from Model~Ia, it seems possible that the comet will evolve
    into an Earth crossing orbit in the future as well as into an
    orbit of larger perihelion distance than 3.0~AU.
    \item Within the time interval [-1000 BC, 4000 AD] the
    aphelion of the comet 67P seems to be under Jupiter's control.
    Outside this time interval the perihelion predominantly
    falls within the zone of Jupiter control.
\end{enumerate}

\section{Acknowledgements}

I am deeply indebted to Professor Grzegorz Sitarski for helpful
discussion and constructive comments on these investigations.

\end{document}